\documentclass[10pt]{article}

\usepackage{amsmath}
\usepackage{amssymb}
\usepackage{amsthm}
\usepackage{enumerate}
\usepackage[all,arc,curve,color,frame]{xy}
\usepackage[utf8]{inputenc}

\usepackage{color, graphicx}
\usepackage[square,sort,comma,numbers]{natbib}

\usepackage{authblk}

\usepackage{geometry}
\usepackage{graphicx} 
\usepackage{amsthm}
\usepackage{amsmath}
\usepackage{amssymb}
\usepackage{booktabs} 
\usepackage{url}

\usepackage{algorithm}      
\usepackage{algpseudocode}  

\usepackage{arydshln} 
\usepackage{tikz}
\usetikzlibrary{arrows.meta}

\renewcommand\section{\@startsection{section}{1}{\z@}
                                   {-3.5ex \@plus -1ex \@minus -.2ex}
                                   {2.3ex \@plus .2ex}
                                   {\normalfont\large\bfseries}}
\renewcommand\subsection{\@startsection{subsection}{2}{\z@}
                                   {-3.25ex\@plus -1ex \@minus -.2ex}
                                   {1.5ex \@plus .2ex}
                                   {\normalfont\normalsize\bfseries}}
\renewcommand\subsubsection{\@startsection{subsubsection}{3}{\z@}
                                   {-3.25ex\@plus -1ex \@minus -.2ex}
                                   {1.5ex \@plus .2ex}
                                   {\normalfont\normalsize\bfseries}}
\renewcommand\paragraph{\@startsection{paragraph}{4}{\z@}
                                   {3.25ex \@plus1ex \@minus.2ex}
                                   {-1em}
                                   {\normalfont\normalsize\bfseries}}

\makeatother

\newdimen\tableauside\tableauside=1.0ex
\newdimen\tableaurule\tableaurule=0.4pt
\newdimen\tableaustep
\def\phantomhrule#1{\hbox{\vbox to0pt{\hrule height\tableaurule
width#1\vss}}}
\def\phantomvrule#1{\vbox{\hbox to0pt{\vrule width\tableaurule
height#1\hss}}}
\def\sqr{\vbox{%
  \phantomhrule\tableaustep

\hbox{\phantomvrule\tableaustep\kern\tableaustep\phantomvrule\tableaustep}%
  \hbox{\vbox{\phantomhrule\tableauside}\kern-\tableaurule}}}
\def\squares#1{\hbox{\count0=#1\noindent\loop\sqr
  \advance\count0 by-1 \ifnum\count0>0\repeat}}
\def\tableau#1{\vcenter{\offinterlineskip
  \tableaustep=\tableauside\advance\tableaustep by-\tableaurule
  \kern\normallineskip\hbox
    {\kern\normallineskip\vbox
      {\gettableau#1 0 }%
     \kern\normallineskip\kern\tableaurule}%
  \kern\normallineskip\kern\tableaurule}}
\def\gettableau#1 {\ifnum#1=0\let\next=\null\else
  \squares{#1}\let\next=\gettableau\fi\next}

\makeatletter

\renewcommand\section{\@startsection{section}{1}{\z@}
                                   {-3.5ex \@plus -1ex \@minus -.2ex}
                                   {2.3ex \@plus .2ex}
                                   {\normalfont\large\bfseries}}
\renewcommand\subsection{\@startsection{subsection}{2}{\z@}
                                   {-3.25ex\@plus -1ex \@minus -.2ex}
                                   {1.5ex \@plus .2ex}
                                   {\normalfont\normalsize\bfseries}}
\renewcommand\subsubsection{\@startsection{subsubsection}{3}{\z@}
                                   {-3.25ex\@plus -1ex \@minus -.2ex}
                                   {1.5ex \@plus .2ex}
                                   {\normalfont\normalsize\bfseries}}
\renewcommand\paragraph{\@startsection{paragraph}{4}{\z@}
                                   {3.25ex \@plus1ex \@minus.2ex}
                                   {-1em}
                                   {\normalfont\normalsize\bfseries}}

\makeatother

\newcommand{\be}{\begin{equation}}
\newcommand{\ee}{\end{equation}}
\newcommand{\bea}{\begin{eqnarray}}
\newcommand{\eea}{\end{eqnarray}}
\newcommand{\ba}{\begin{array}}
\newcommand{\ea}{\end{array}}

\newcommand{\id}{\hbox{1\kern-.27em l}}

\newcommand{\emp}{\emptyset}

\newtheorem{prop}{Proposition}[section]

\newtheorem{remark}{Remark}[section]

\begin{document}

\title{Exact $S$-duality Map for Rigid Surface Operators}



\author[1]{Chuanzhong Li}
\author[1]{Xiaoman Luo}
\author[2]{Bao Shou}
\affil[1]{College of Mathematics and Systems Science, Shandong University of Science and Technology, Qingdao, 266590, P.R. China}

\affil[2]{Center of Mathematical Sciences, Zhejiang University, Hangzhou 310027, China}

\affil[ ]{\textit{E-mail:} \texttt{lichuanzhong@sdust.edu.cn}\quad \texttt{$luoxiaoman\_yjs$@163.com}\quad\texttt{bsoul@zju.edu.cn}} 
\date{}
\maketitle

\begin{abstract}

Surface operators in four-dimensional gauge theories are two-dimensional defects, serving as natural generalizations of Wilson lines and 't Hooft line operators. 
They act as ideal probes for exploring the non-perturbative structure of the theory. Rigid surface operators are a specific class of surface operators characterized by the absence of continuous deformation parameters. It is expected that a closed $S$-duality map should exist among these rigid operators. 
While progress has been made on specific examples or subclasses by leveraging invariants and empirical conjectures, a complete picture remains elusive.

A significant challenge arises when multiple rigid surface operators share identical invariants, making the determination of $S$-duality relations difficult. More critically, a mismatch exists in the number of rigid surface operators between dual theories when classified by invariants; this is referred to as the \textit{mismatch problem}. 
This discrepancy suggests the necessity of extending the scope of consideration beyond strictly rigid operators.
 In this paper, we propose a direct, natural, and precise $S$-duality map for rigid surface operators.
  Our map is realized by moving the longest row in the pair of partitions defining a surface operator from one factor to the other, with an additional box appended or deleted to balance the total number of boxes. 
  This mapping naturally incorporates non-rigid surface operators, thereby resolving the mismatch problem. The proposed map is applicable to gauge groups of all ranks and clarifies several long-standing puzzles in the field.

\end{abstract}


\paragraph{Introduction}

In four-dimensional gauge theories, surface operators are a class of non-local observables supported on a two-dimensional surface $(x^0,x^1) \subset \mathbb{R}^4$. They constitute natural generalizations of Wilson lines and 't Hooft line operators \cite{GW06, GW08, nr}. 
Investigating the $S$-duality of surface operators provides profound insights into the non-perturbative structure of the theory \cite{defect}. $S$-duality maps a theory with gauge group $G$ and coupling constant $\tau$ to a theory with the Langlands dual group $^LG$ and coupling $-1/n_{\mathfrak{g}}\tau$, where $n_{\mathfrak{g}}$ is a group-dependent constant. 
Some canonical examples of $S$-duality pairs are listed as follows \cite{Wy09}:
\begin{equation*}
	\begin{array}{l l l}
		G & ^LG & \mathbb{Z}(G) \\
		\hline
		\mathrm{Spin}(2n+1) & \mathrm{Sp}(2n)/\mathbb{Z}_2 & \mathbb{Z}_2 \\
		\mathrm{Sp}(2n) & \mathrm{Spin}(2n+1)/\mathbb{Z}_2 \equiv \mathrm{SO}(2n+1) & \mathbb{Z}_2 \\
		\mathrm{SO}(2n) & \mathrm{SO}(2n) & \mathbb{Z}_2 \\
	\end{array}
\end{equation*}
Here, $\mathbb{Z}(G)$ denotes the center of the group $G$. 

The pioneering work of Gukov and Witten \cite{GW06} constructed a class of half-BPS surface operators depending on continuous parameters. In subsequent work, they focused on \textit{rigid surface operators}, which do not depend on any continuous parameters. It is naturally expected that the set of rigid surface operators is closed under $S$-duality. 
By utilizing invariants of rigid surface operators under $S$-duality, and relying on case-by-case conjectures, $S$-duality relations have been established for simple examples and specific subclasses.

However, a mismatch in the number of rigid surface operators with identical invariants between the $B_n$ and $C_n$ theories was identified in \cite{GW08, Wy09}. Fortunately, in our previous work, we clarified the origin of this discrepancy, identifying the root cause within the rigidity conditions themselves \cite{rso}.
 Furthermore, we significantly simplified the calculation of invariants and clarified their interrelationships, laying the foundation for resolving these issues \cite{Shou-sc, sl}.

In this paper, we propose a precise and self-consistent $S$-duality map that holds for gauge groups of all ranks, by extending the scope to include certain non-rigid surface operators. The $S$-duality map is realized simply by moving rows within the partitions representing the surface operators. 
This approach not only reproduces existing results but also reveals structure beyond initial expectations. 
The proposed map adheres to Occam's Razor in its simplicity. We also provide a characterization of the non-rigid surface operators introduced via this $S$-duality map.

\paragraph{Rigid Surface Operators}

In $\mathcal{N}=4$, Super Yang-Mills (SYM) theory, half-BPS surface operators supported on $(x^0,x^1) \subset \mathbb{R}^4$ require the gauge field components $A$ and scalar field components $\phi$ normal to the surface to satisfy the Hitchin equations \cite{GW08}:
\begin{equation}\label{hitch}
	F_A - \phi \wedge \phi = 0,\quad  D_A\phi =0,\quad D_A\star A = 0 \,.
\end{equation}
A surface operator is defined as a solution to these equations with a prescribed singularity along the surface $(x^0,x^1)$. Setting $x_2+ix_3 = re^{i\theta}$, the most general rotation-invariant ansatz for $A$ and $\phi$ is:
\begin{equation*}
	A = a(r) \, d\theta \,, \quad \phi = -c(r) \, d\theta + b(r) \frac{dr}{r} \,.
\end{equation*}
Substituting this ansatz into the Hitchin equations (\ref{hitch}) and defining $s = -\ln r$, the equations reduce to Nahm's equations:
\begin{eqnarray} \label{nahm}
	\frac{da}{ds} &=& [b,c]\,, \nonumber \\
	\frac{db}{ds} &=& [c,a] \,,\\
	\frac{dc}{ds} &=& [a,b] \, .\nonumber
\end{eqnarray}
For example, with the gauge group $G=SU(2)$, these equations are solved by:
\begin{equation} \label{nahmshou}
	a = \frac{t_x}{s + 1/f}\,,\qquad b = \frac{t_z}{s + 1/f}\,,\qquad c = \frac{t_y}{s + 1/f} \,,
\end{equation}
where $t_x, t_y$, and $t_z$ span an adjoint representation of $\mathfrak{su}(2)$. The surface operator is conformally invariant if the function $f$ is allowed to fluctuate.

Alternatively, surface operators can be characterized by the conjugacy class of the monodromy:
\begin{equation}
	U = P \exp\left(\oint \mathcal{A}\right) \,,
\end{equation}
where $\mathcal{A} = A + i \phi$ is the solution to the flatness condition $\mathcal{F} = d\mathcal{A} + \mathcal{A}\wedge \mathcal{A}=0$ following from (\ref{hitch}). $U$ is independent of the integration contour around the singularity at $r=0$. For the surface operators in (\ref{nahmshou}), $U$ takes the form:
\begin{equation} \label{Uplusshou}
	U= P \exp\left(\frac{2\pi}{s+1/f} \,\,  t_+ \right) \,,
\end{equation}
where $t_+\equiv t_x +i t_y$ is nilpotent (i.e., $t_+^{n}=0$ for some positive integer $n$). Consequently, the surface operator corresponding to the group element $U$ is called a \textit{unipotent surface operator}. For a general group $G$, $t_+$ in formula (\ref{Uplusshou}) can be described in a block-diagonal basis as:
\begin{equation}
	\label{ti}
	t_+ = \left( \begin{array}{ccc} t_+^{\lambda_1}  & & \\
		& \ddots &   \\
		& & t_+^{\lambda_l}
	\end{array} \right ),
\end{equation}
where the notation $t_+^{\lambda_k}$ stands for the `raising' generator of the $\lambda_k$-dimensional irreducible representation of $\mathfrak{su}(2)$. 

\begin{figure}[!ht]
	\begin{center}
		\includegraphics[width=5in]{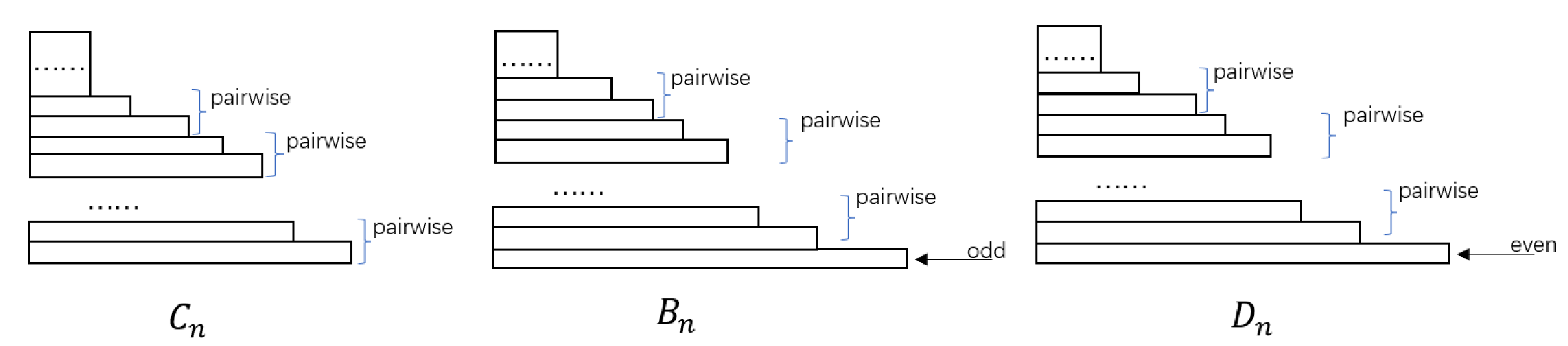}
	\end{center}
	\caption{Partitions with $2n+1$, $2n$, and $2n$ boxes in the $B_n$, $C_n$, and $D_n$ theories, respectively.}
	\label{bdcd}
\end{figure}

From the above, a general unipotent surface operator can be characterized by a partition $\lambda_1^{n_1}\lambda_{2}^{n_{2}}\cdots\lambda_m^{n_m}$, where $\lambda_i \in \mathbb{N}$ with $\lambda_1> \lambda_2 > \cdots > \lambda_m$, and $n_k$ denotes the multiplicity of $t_+^{\lambda_k}$ in the matrix (\ref{ti}) \cite{Wy09}. A partition corresponds to a Young tableau. For example:
\begin{equation}
	\lambda=3^22^31^2:\quad	\tableau{2 5 7}
\end{equation}
The restricted partitions for unipotent surface operators are illustrated in Fig.(\ref{bdcd}) \cite{Wy09}. 
\begin{itemize}
	\item For a rigid $B_n$ partition, the longest row always contains an odd number of boxes. The subsequent rows appear in pairs, either both odd or both even in length.
	\item For a rigid $D_n$ partition, the longest row always contains an even number of boxes, followed by a pairwise pattern.
	\item For a rigid $C_n$ partition, the two longest rows both contain either an even or an odd number of boxes, followed by a pairwise pattern.
\end{itemize}

A unipotent conjugacy class (surface operator) is called \textit{rigid} if its dimension is strictly smaller than that of any nearby orbit. The corresponding partitions must satisfy the \textit{rigid conditions}:
\begin{enumerate}
	\item No gaps, i.e., $\lambda_i-\lambda_{i+1}\leq1$ for all $i$.
	\item No odd (even) part appears exactly twice in a partition of the $B_n$ or $D_n$ ($C_n$) theories \cite{GW08}.
\end{enumerate}
Rigidity implies the absence of continuous deformation parameters, corresponding to an isolated conjugacy orbit.

A \textit{semisimple class} is another type of conjugacy class in a Lie group that can lead to surface operators. Given a semisimple element $S$ (a diagonal matrix), such operators are defined by imposing twisted boundary conditions around the surface: $\Psi(r, \theta+2\pi) = S \Psi(r, \theta) S^{-1}$. If the surface operator corresponding to $S$ is rigid, then $S$ has the form $ S=\text{diag}( {+1, \dots, +1}, \underbrace{-1, \dots, -1}_{2i \text{ times}} ) $. Its centralizer $G_S$ has maximal local dimension, and its Lie algebra is broken from the gauge group $G$ as follows \cite{Wy09}:
\begin{align}\label{lie}
	\mathfrak{so}(2n+1) &\to \mathfrak{so}(2k+1) \oplus \mathfrak{so}(2n - 2k), \nonumber\\
	\mathfrak{sp}(2n) &\to \mathfrak{sp}(2k) \oplus \mathfrak{sp}(2n - 2k), \\
	\mathfrak{so}(2n) &\to \mathfrak{so}(2k) \oplus \mathfrak{so}(2n - 2k). \nonumber
\end{align}
In summary, the most general rigid surface operator is given by a single-valued group element:
\begin{equation}\label{vv}
	V=SU, 
\end{equation}
where $S$ is a rigid semisimple element, and $U$ is a rigid unipotent element within the centralizer $G_S$. Corresponding to formula (\ref{lie}), the general rigid surface operators in different theories are given by partition pairs:
\begin{equation}\label{rsp}
	(\lambda'_{B_k},\lambda''_{D_{n-k}})_{B_n}, \quad (\lambda'_{C_k},\lambda''_{C_{n-k}})_{C_n}, \quad (\lambda'_{D_k},\lambda''_{D_{n-k}})_{D_n}. 
\end{equation}
A unipotent surface operator corresponds to the case where one of the partitions is empty. Existing work indicates that the number of rigid surface operators differs on the two sides of $S$-duality. Therefore, to establish a one-to-one $S$-duality relationship, non-rigid surface operators must be introduced. Our basis for extending this scope lies in the invariants of surface operators under $S$-duality.

\paragraph{Invariants}

Quantities that remain invariant under $S$-duality transformations include the Dimension ($Dim$), Fingerprint ($FP$), Symbol ($Sym$), and discrete invariants (Center and Topology). The original definitions of these invariants are often complex. Fortunately, we have previously demonstrated \cite{sl, rso} that:
\begin{equation}\label{fpd}
	Sym \sim FP \succcurlyeq Dim, 
\end{equation}
meaning the first two are equivalent, and both are finer than $Dim$. Therefore, we will focus exclusively on the discussion of the symbol invariant.

\begin{table}[t]
	\centering 
	\begin{tabular}{|c|c|c|c|}\hline
		Length & Parity  & Location in pairwise rows   & Contribution    \\ \hline
		$2n+1$ & odd & top & $\Bigg(\!\!\!\ba{c}0 \;\; 0\cdots 0\;\; 0 \cdots 0 \\
		\;\;\;0\cdots \underbrace{1 \;\;1\cdots 1}_{n} \ \ea \Bigg)$   \\ \hline
		$2n+1$ & odd & bottom   & $\Bigg(\!\!\!\ba{c}0 \;\; 0\cdots \overbrace{ 1\;\; 1\cdots1}^{n+1} \\
		\;\;\;0\cdots 0\;\; 0\cdots 0 \ \ea \Bigg)$    \\ \hline
		$2m$ & even & bottom   &  $\Bigg(\!\!\!\ba{c}0 \;\; 0\cdots 0\;\; 0 \cdots 0 \\
		\;\;\;0\cdots \underbrace{1 \;\;1\cdots 1}_{m} \ \ea \Bigg)$  \\ \hline
		$2m$ & even & top    &   $\Bigg(\!\!\!\ba{c}0 \;\; 0\cdots \overbrace{ 1\;\; 1\cdots1}^{m} \\
		\;\;\;0\cdots 0\;\; 0\cdots 0 \ \ea \Bigg)$      \\ \hline
	\end{tabular}
	\caption{Contributions of rows in partitions to the symbol. The first two columns indicate the length and parity of the rows.}
	\label{st}
\end{table}

In \cite{Shou-sc}, we provided a construction of the symbol invariant that is independent of the specific theory \cite{rso}, as shown in Table \ref{st}. Note that the first row in a $B_n$ or $D_n$ partition is considered the top row of a pairwise set, as proved in \cite{Shou-sc}. The partition-row contributions to the symbol remain invariant under two operations: direct movement between the same positions in pairwise rows; and movement between different positions in pairwise rows accompanied by adding or subtracting a box at the end of the row.

The symbol $\sigma$ of the surface operators in formula (\ref{rsp}) is obtained by adding the entries that are `in the same place' from the symbols of $\lambda'$ and $\lambda''$:
\begin{equation}\label{ddddr}
	\sigma((\lambda';\lambda''))=\sigma(\lambda')+\sigma(\lambda'').
\end{equation}
We illustrate the calculation details using $(1,2^21^4)_B$ as an example:
\begin{equation*} \label{symboladd}
	\begin{aligned}
		\sigma((1,2^21^4)_B) &= \sigma(\tableau{1})_B+\sigma(\tableau{2 6})_D \\
		&= \left(\begin{array}{@{}c@{}c@{}c@{}c@{}c@{}c@{}c@{}c@{}c@{}c@{}c@{}c@{}c@{}} 0 \\  & \end{array} \right)_B+\left\{ 	\left(\begin{array}{@{}c@{}c@{}c@{}c@{}c@{}c@{}c@{}c@{}c@{}c@{}c@{}c@{}c@{}} 1&&1&&1 \\ &0&&0 & \end{array} \right) +
		\left(\begin{array}{@{}c@{}c@{}c@{}c@{}c@{}c@{}c@{}c@{}c@{}c@{}c@{}} 0&&0&&0 \\ &0&&1 & \end{array} \right)\right\}_D \\
		&= \left(\begin{array}{@{}c@{}c@{}c@{}c@{}c@{}c@{}c@{}c@{}c@{}c@{}c@{}c@{}c@{}} 1&&1&&1 \\ &0&&1 & \end{array} \right).
	\end{aligned}
\end{equation*}
First, according to formula (\ref{ddddr}), we calculate the symbol invariant for each partition separately. Second, according to Table \ref{st}, we sum the contributions of each row within each partition. The addition is performed entry-wise, aligning the rows to the right and padding with zeros if necessary. According to Table \ref{st}, the contribution of the first partition to the symbol invariant is zero.

$S$-duality implies a Center vs. Topology duality, typically manifested as $\mathbb{Z}(G) \leftrightarrow \pi_1(^LG_{\text{ad}})$, where $\pi_1(^LG_{\text{ad}})$ is the fundamental group of the adjoint form of the dual group $^LG$ \cite{GW06, GW08}. From Table 1, it is evident that the theories we are concerned with have only one non-trivial center element $\zeta =-1$. For $\zeta  \in \mathbb{Z}(G)$, if there does not exist $g\in G$ such that the surface operator $V$ in formula (\ref{vv}) satisfies:
\begin{equation}\label{dc}
	gVg^{-1}=\zeta V,
\end{equation} 
then the surface operator $V$ is said to \textit{detect the center}. In this case, $S$-duality requires that the dual surface operator $^LV$ must be able to \textit{detect the topology}. This is equivalent to proving that the homomorphism $\iota: \pi_1(H) \rightarrow \pi_1(^LG)$ induced by the inclusion map $\iota: H\rightarrow ^LG$ is surjective, where $H$ is the symmetry group of $^LV$ \cite{GW08}. Conversely, if one side detects the topology, the dual theory must detect the center.

The Center-Topology duality serves as a consistency check for our constructed $S$-duality map. A unipotent surface operator $V$ always detects the center. We know that the eigenvalues of $V$ are always 1 in any representation of the group, whereas the eigenvalues of $\zeta V$ depend on the value of $\zeta$ in that specific representation; thus, no conjugacy relation exists as in formula (\ref{dc}).

\paragraph{$S$-duality Map}

\begin{figure}[!ht]
	\begin{center}
		\includegraphics[width=6in]{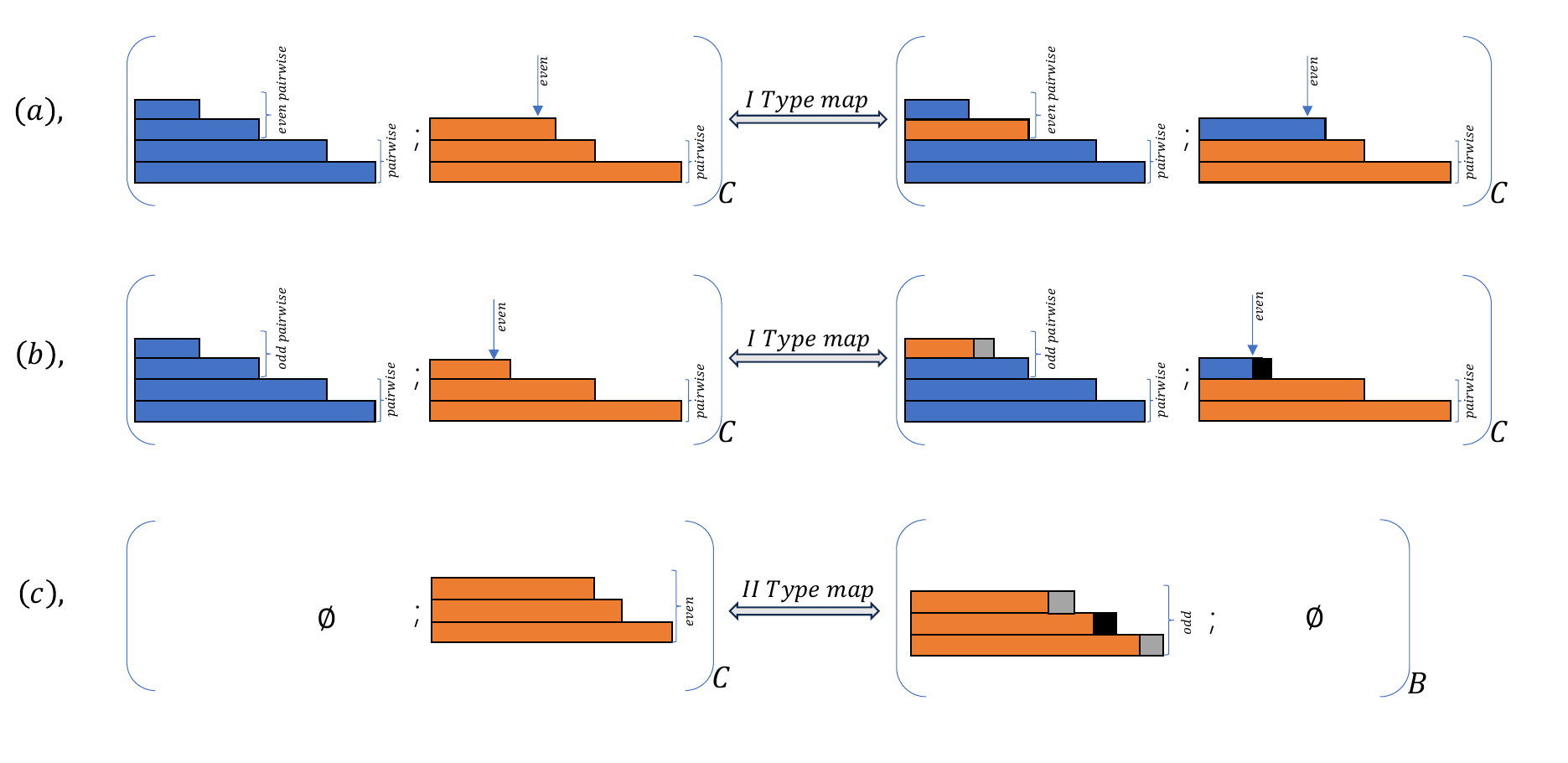}
	\end{center}
	\caption{Symbol preserving row moves: Type $I$ and Type $II$ maps. Black boxes indicate deletion at the end of rows, while gray boxes indicate appending.}
	\label{tm}
\end{figure}

Fig.(\ref{tm}) illustrates two classes of symbol-preserving maps \cite{rso, sl}: Type $I$ maps transform surface operators within the same theory, while Type $II$ maps transform surface operators to a different theory, consistent with Table \ref{st}.
\begin{itemize}
	\item[(a)] Two rows belonging to different factors of a surface operator swap their positions while remaining in the same locations within the pairwise structure.
	\item[(b)] The swapped rows are in different locations within the pairwise structure; appending or removing one box preserves the symbol.
	\item[(c)] A $C_n$ unipotent operator with only even rows and a $B_n$ unipotent operator with only odd rows map into each other.
\end{itemize}

The $S$-duality map between $B_n$ and $C_n$ theories belongs to the Type $II$ class.

\begin{figure}[!ht]
	\begin{center}
		\includegraphics[width=6in]{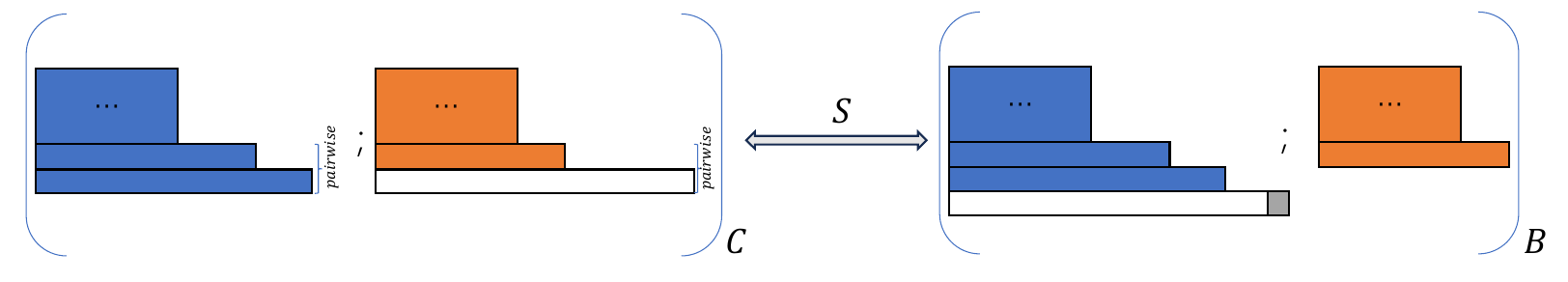}
	\end{center}
	\caption{$S$-duality map: the white longest row is moved from one partition to the other within a rigid surface operator. The gray box is appended at the end of the row.}
	\label{sd}
\end{figure}

\begin{prop}\label{sm}
	The $S$-duality map is realized by moving the longest row among the two partitions of a rigid surface operator from one partition to the other, as shown in Fig. \ref{sd}. For the mapping $C_n \rightarrow B_n$, the total number of boxes increases by 1, so a box is appended; for $B_n\rightarrow C_n$, a box is deleted.
\end{prop}

\begin{figure}[!ht]
	\begin{center}
		\includegraphics[width=6in]{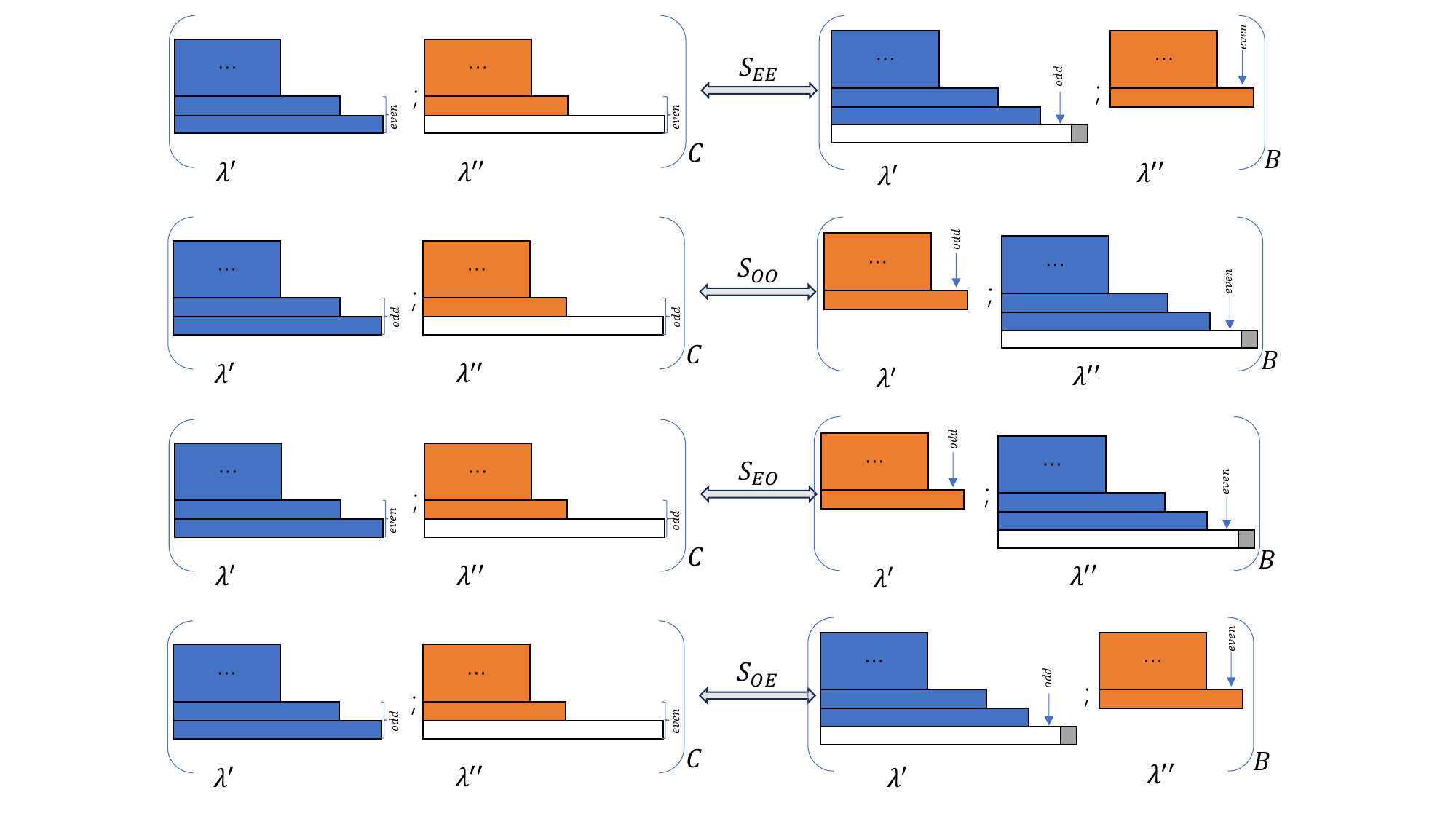}
	\end{center}
	\caption{Four cases of the $S$-duality map: $(\lambda'_{C_k},\lambda''_{C_{n-k}})_{C_n}  \leftrightarrow     (\lambda'_{B_k},\lambda''_{D_{n-k}})_{B_n}$.}
	\label{s4}
\end{figure}

The proposed $S$-duality map can be refined into four cases, as depicted in Fig.(\ref{s4}). Without loss of generality, assume that the first row of the partition $\lambda''$ is the longest for the rigid surface operator $(\lambda', \lambda'')$ in the $C_n$ theory (we adopt the same convention when the first rows in the two factors have equal length).
For the dual maps $S_{EE}$ and $S_{OO}$, the first two rows of both factors in the $C_n$ surface operator share the same parity. For $S_{EO}$ and $S_{OE}$, if the longest row lies in the second factor of the $B_n$ surface operator, the map is $S_{EO}$; otherwise, it is $S_{OE}$.

\paragraph{Evidences}
According to Fig.(\ref{s4}), the number of boxes in the surface operator in the $B_n$ theory is one greater than that in the $C_n$ theory, as expected. Based on Table \ref{st}, the symbol invariants are preserved under the proposed $S$-duality map. Furthermore, the contribution to the symbol invariant from the longest row is distinctively the longest. Thus, for a generic surface operator, the longest row can only be moved in the manner prescribed in Fig.(\ref{s4}).

\begin{figure}[!ht]
	\begin{center}
		\includegraphics[width=6in]{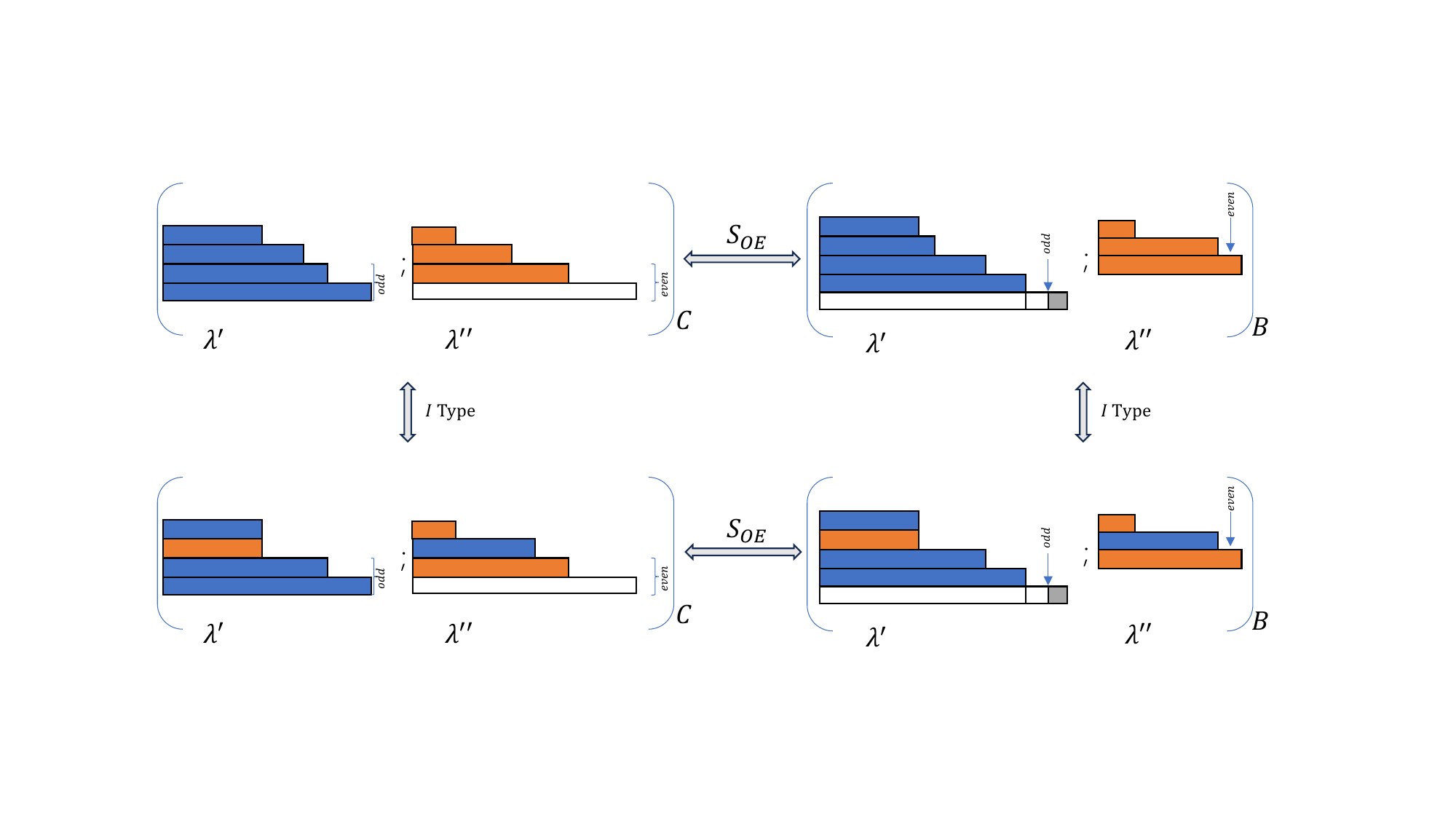}
	\end{center}
	\caption{Type $I$ maps are operations that preserve the symbol invariant within the same theory. The operator in the bottom left violates Rigid Condition 1 (Gap of 2). The operator in the top right violates Rigid Condition 2 (Part 1 appears twice). The operator in the bottom right violates both conditions.}
	\label{ws}
\end{figure}

Non-rigid surface operators emerge naturally within our proposed $S$-duality map, as shown in Fig.(\ref{ws}). The first $S_{OE}$ map realizes an $S$-duality between a rigid surface operator and a non-rigid surface operator. The downward arrow on the left transforms a rigid surface operator into a non-rigid one. Notably, the second $S_{OE}$ in this figure realizes a duality between two non-rigid surface operators. According to 
Fig.(\ref{ws}), the proposed $S$-duality map not only relates operators between dual theories but also intertwines with local moves on one side to corresponding moves on the other, thereby ensuring that the total number of surface operators (including non-rigid ones) matches on both sides, as anticipated.

\begin{table}[h!]
	\centering
	\small
	\begin{tabular}{l | l l l : l l l l l}
		{} &
		{$B_{4}$} &
		{$Map$} &
		{$C_{4}$} &
		{$B_{4}$} &
		{$C_{4}$} &
		{Sym} &
		{FP} &
		{Dim} \\
		\hline
		1 &
		$(1;1^8)$ &  $S_{EO}$
		& 
		$(2\,1^{6};\emptyset)$ &
		$(1;1^8)$ & %
		$(2\,1^{6};\emptyset)$ & %
		$\left(\begin{array}{@{}c@{}c@{}c@{}c@{}c@{}c@{}c@{}c@{}c@{}c@{}c@{}c@{}c@{}c@{}}
			1&&1&&1&&1 \\
			&0&&0&&0&
		\end{array}\right)$ &
		$[1^3;1]$ &
		8 \\
		2 &
		$(2^{2}\,1^{5};\emptyset)$ &   $S_{EE}$
		& 
		$(1^{2};1^{6})$ &
		$(2^{2}\,1^{5};\emptyset)$ &
		$(1^{2};1^{6})$ &
		$\left(\begin{array}{@{}c@{}c@{}c@{}c@{}c@{}c@{}c@{}c@{}c@{}c@{}c@{}c@{}c@{}c@{}}
			0&&0&&0&&0 \\
			&1&&1&&2&
		\end{array}\right)$ &
		$[2\,1^2;\emptyset]$ &
		12 \\
		3 &
		$(2^4\,1;\emptyset)$ &   $S_{EE}$
		& 
		$(1^{4};1^4)$ &
		$(2^4\,1;\emptyset)$ &
		$(1^{4};1^4)$ &
		$\left(\begin{array}{@{}c@{}c@{}c@{}c@{}c@{}c@{}c@{}c@{}c@{}c@{}c@{}c@{}c@{}c@{}}
			0&&0&&0 \\
			&2&&2&
		\end{array}\right)$ &
		$[2^2;\emptyset]$ &
		16 \\
		$4^{*}$ &
		${(1^{3};1^6)}$ &  $S_{EO}$
		&  
		${(2^{3}\,1^2;\emptyset)}$ &
		${(1^{3};1^6)}$ &
		${(2^{3}\,1^2;\emptyset)}$ &
		$\left(\begin{array}{@{}c@{}c@{}c@{}c@{}c@{}c@{}c@{}c@{}c@{}c@{}c@{}c@{}c@{}c@{}}
			1&&1&&1 \\
			&0&&1&
		\end{array}\right)$ &
		$[1;1^3]$ &
		18 \\
		$5^{*}$ &
		${(1;2^2\,1^4)}$ &  $S_{EO}$
		& 
		${(1^{2};2\,1^4)}$ &
		${(1;2^2\,1^4)}$ &
		${(1^{2};2\,1^4)}$ &
		$\left(\begin{array}{@{}c@{}c@{}c@{}c@{}c@{}c@{}c@{}c@{}c@{}c@{}c@{}c@{}c@{}c@{}}
			1&&1&&1 \\
			&0&&1&
		\end{array}\right)$ &
		$[1;1^3]$ &
		18 \\
		6 &
		$(1^{5};1^{4})$ &   $S_{EE}$
		& 
		$(2^4;\emptyset)$ &
		$(1^{5};1^{4})$ &
		$(2\,1^2;1^4)$ &
		$\left(\begin{array}{@{}c@{}c@{}c@{}c@{}c@{}c@{}c@{}c@{}c@{}c@{}c@{}c@{}c@{}c@{}}
			0&&1&&1 \\
			&1&&1&
		\end{array}\right)$ &
		$[\emptyset;1^4]$ &
		20 \\
		7 &
		$(3\,2^{2}\,1^{2};\emptyset)$ &    $S_{OE}$
		& 
		$(2\,1^2;1^4)$ &
		$(1^{5};1^{4})$ &
		$(2\,1^2;1^4)$ &
		$\left(\begin{array}{@{}c@{}c@{}c@{}c@{}c@{}c@{}c@{}c@{}c@{}c@{}c@{}c@{}c@{}c@{}}
			0&&1&&1 \\
			&1&&1&
		\end{array}\right)$ &
		$[\emptyset;1^4]$ &
		20 \\
		8 &
		$(1;3\,2^2\,1)$ &  $S_{OO}$
		& 
		$(2\,1^2;2\,1^2)$ &
		$(1;3\,2^2\,1)$ &
		$(2\,1^2;2\,1^2)$ &
		$\left(\begin{array}{@{}c@{}c@{}c@{}c@{}c@{}c@{}c@{}c@{}c@{}c@{}c@{}c@{}c@{}c@{}}
			2&&2 \\
			&0&
		\end{array}\right)$ &
		$[2;2]$ &
		24 \\
		$9^{*}$ &
		$(2^2\,1;1^{4})$ &   $S_{EO}$
		& 
		$(3^22;\emptyset)$ &
		$(2^2\,1;1^{4})$ &
 	\quad	\quad $\bigtriangleup$ &
		$\left(\begin{array}{@{}c@{}c@{}c@{}c@{}c@{}c@{}c@{}c@{}c@{}c@{}c@{}c@{}c@{}c@{}}
			1&&1 \\
			&2&
		\end{array}\right)$ &
		$[3;1]$ &
		24 \\
		\hline
		{} &
		{$B_{4}$} &
		{$Map$} &
		{$C_{4}$} &
		{$B_{4}$} &
		{$C_{4}$} &
		{Sym} &
		{FP} &
		{Dim} \\
		\hline
		10 &
		$(1;1^{2N})$ &   $S_{EO}$
		& 
		$(\emptyset;21^{2N-2})$ &
		$(1;1^{2N})$ &$(\emptyset;21^{2N-2})$ &
		$\left(\begin{array}{@{}c@{}c@{}c@{}c@{}c@{}c@{}c@{}c@{}c@{}c@{}c@{}c@{}c@{}c@{}}
			1&&\cdots&&1&&1 \\
			&0&&\cdots&&0&
		\end{array}\right)$ &
		$[1^{2N-5};1]$ &
		2N \\
		11 &
		$(2^{2}\,1^{2N-3};\emptyset)$ &  $S_{EE}$
		& 
		$(1^{2};1^{2N-2})$ &
		$(2^{2}\,1^{2N-3};\emptyset)$ &
		$(1^{2};1^{2N-2})$ &
		$\left(\begin{array}{@{}c@{}c@{}c@{}c@{}c@{}c@{}c@{}c@{}c@{}c@{}c@{}c@{}c@{}c@{}}
			0&&\cdots&&0&&0 \\
			&1&&\cdots&&2&
		\end{array}\right)$ &
		$[2\,1^{2N-6};\emptyset]$ &
		3N \\
		$12^{*}$ &
		$\textcolor{red}{(1^{2N-3};1^4)}$ &   $S_{EE}$
		& 
		$(2^{2N-4};\emptyset)$ &
		$\textcolor{red}{(1^{2N-3};1^4)}$ &
		$\textcolor{blue}{(21^2;1^{2N-4})}$ &
		$\left(\begin{array}{@{}c@{}c@{}c@{}c@{}c@{}c@{}c@{}c@{}c@{}c@{}c@{}c@{}c@{}c@{}}
			0&&\cdots&&1&&1 \\
			&0&&\cdots&&1&
		\end{array}\right)$ &
		$[\emptyset;1^{2N-4}]$ &
		5N \\
		$13^{*}$ &
		$(32^21^{2N-6}; \emptyset)$ &  $S_{OE}$
		&  
		$\textcolor{blue}{(21^2;1^{2N-4})}$ &
		--- &
		--- &
		$\left(\begin{array}{@{}c@{}c@{}c@{}c@{}c@{}c@{}c@{}c@{}c@{}c@{}c@{}c@{}c@{}c@{}}
			0&&\cdots&&1&&1 \\
			&0&&\cdots&&1&
		\end{array}\right)$ &
		$[\emptyset;1^{2N-4}]$ &
		5N \\
	\end{tabular}
	\caption{The first three columns show the dual pair of surface operators and their specific $S$-duality map given by Proposition \ref{sm}. Columns 4 and 5 show the dual pairs given by \cite{GW08}. Triangle/Red/Blue indicates discrepancies between two proposals.}
	\label{gwt}
\end{table}

The proposed $S$-duality map perfectly reproduces existing results while resolving long-standing puzzles. Taking $G=SO(9)$ and $^LG=Sp(8)$ as an example (shown in Table \ref{gwt}), we observe that the symbol and fingerprint invariants of different surface operators are consistent,
 and identical symbols correspond to the same $Dim$. This confirms that the symbol and fingerprint are equivalent invariants, and the symbol is finer than $Dim$. In Examples 1, 2, 3, and 6, the $S$-duality pairs can be determined solely by invariants, as each has a unique candidate.
  These pairs correspond exactly to the one-to-one mapping provided by our $S$-duality map.

Unfortunately, the entries marked with an asterisk in the table are examples where $S$-duality was not successfully realized in \cite{GW08}. In Examples 5 and 6 with $Dim=18$, the operators on both sides of the theory share identical invariants, making it impossible to determine the $S$-duality pairs based on invariants alone. 
Our $S$-duality map resolves this by achieving a one-to-one dual match.
A more serious issue arises in Examples 8 and 9 with $Dim=24$, where there are two operators on the $B_4$ side but only one on the $C_4$ side. This is a quintessential example of the mismatch problem. 
In Example 9, using Proposition \ref{sm}, $(2^21;1^4)$ finds its dual operator $(3^22;\emptyset)$, which is, however, a non-rigid surface operator with a gap in the last part.

Examples 10 to 13 in Table \ref{gwt} are dual pairs that involve rigid semisimple conjugacy classes on at least one side. In the last two examples, the pairs marked in red and green are the duality results given by Gukov and Witten \cite{GW08}, who noted a discrepancy with the results of \cite{l8}. However, the $S$-dual pairs obtained through our construction are consistent with \cite{l8}. Furthermore, our $S$-duality map perfectly and self-consistently reproduces other examples from  \cite{GW08}, presented in supplementary materials \cite{sss}.

Additionally, for the subclass of rigid surface operators of the form $(\rho, \rho)$, the proposed $S$-duality map by Wyllard in \cite{Wy09} aligns with ours, although derived differently. For all these examples, the consistency of discrete invariants has been verified \cite{GW08, Wy09}.

\paragraph{Special Rigid Unipotent Surface Operators}
Special rigid surface operators are those whose partition rows all have consistent parity, denoted by $\rho_{even}$ or $\rho_{odd}$. For this type, Gukov and Witten proposed the following $S$-duality \cite{GW08}:
\begin{equation}\label{special}
	(\emptyset, \rho_{even})_C\longleftrightarrow (\rho'_{odd},\emptyset)_B. \quad \quad \text{Example}: (  \tableau{4 6},\emptyset)_C\longleftrightarrow (\tableau{1 3 7},\emptyset)_B.
\end{equation}
They verified that this duality preserves all invariants.

\begin{figure}[!ht]
	\begin{center}
		\includegraphics[width=5in]{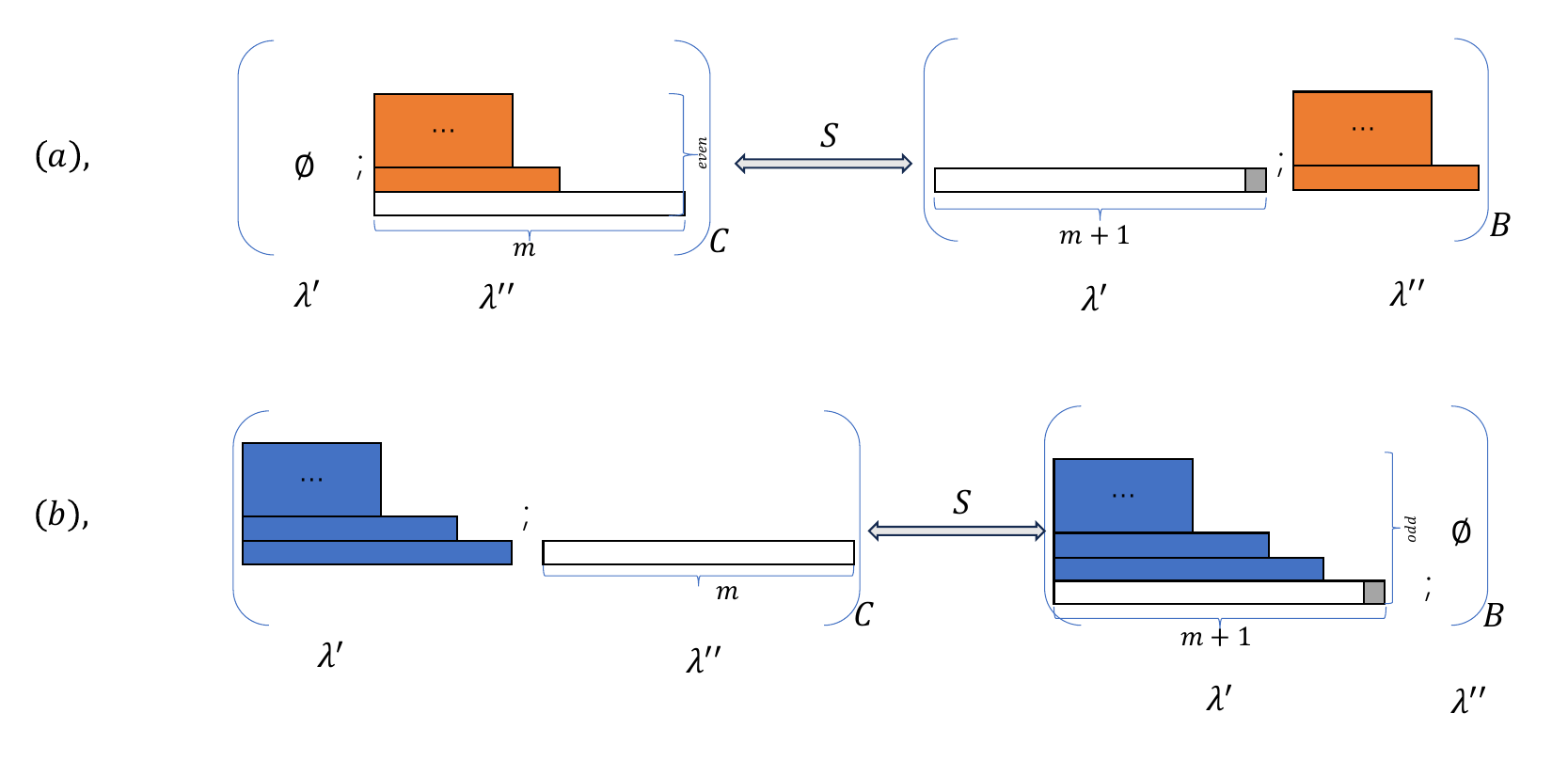}
	\end{center}
	\caption{$S$-duality map for rigid special unipotent surface operators given by Proposition \ref{sm}.}
	\label{sr}
\end{figure}

According to Proposition \ref{sm}, the dual surface operators for special rigid unipotent surface operators are $(1^{m+1}, \lambda'')_{B_{n}}$ and $(\lambda', 1^{m} )_{C_{n}}$ as shown in Fig.(\ref{sr}), which differs significantly from formula (\ref{special}). Since our map naturally preserves the symbol, we only need to verify the self-consistency of discrete invariants under $S$-duality for this subclass.

We already know that a unipotent surface operator always detects the center. Therefore, we must prove that its dual operators, $(1^{m+1}, \lambda'')_{B_{n}}$ and $(\lambda', 1^{m} )_{C_{n}}$, are capable of detecting the topology. The factors $1^{m+1}$ and $1^{m}$ in these operators ensure that the corresponding operators retain sufficient symmetry $H$. Consequently, we can always guarantee that the natural map $\iota: \pi_1(H) \rightarrow \pi_1(^LG)$ is surjective.

In Fig.(\ref{sr})(a), the surface operator guarantees at least an $SO(2)$ symmetry (since the odd part 1 has a multiplicity of at least 2). The map $\iota: \pi_1(SO(2)) \rightarrow \pi_1(SO(2n+1))$ is surjective. Therefore, $(1^{m+1}, \lambda'')_{B_{n}}$ detects the topology.
In Fig.(\ref{sr})(b), for $(\lambda', 1^{m} )_{C_{n}}$, the part 1 appears $m$ times in the second factor, preserving at least an $Sp(m)$ symmetry. Since $m$ is even, the central element $-1$ and $1$ in $Sp(m)$ can be connected to form a generator of the fundamental group of $Sp(2n)/\mathbb{Z}_2$. Thus, we obtain a surjective natural map $\iota: \pi_1(Sp(m)) \rightarrow \pi_1(Sp(2n)/\mathbb{Z}_2)$, meaning $(\lambda', 1^{m} )_{C_{n}}$ detects the topology.

In summary, although our proposed $S$-duality map differs from that of \cite{GW08}, it successfully passes the check for discrete invariants.\footnote{Conjecture: The symbol invariant and its construction completely determine the rigid surface operators. Thus, the symbol implies the discrete invariants.}
It is not surprising that discrete invariants remain unchanged under different $S$-duality mapping schemes for rigid surface operators, as these invariants are relatively coarse. Following the logic of formula (\ref{special}), a more general $S$-duality map is proposed  by Wyllard \cite{Wy09}:
$$(\rho_{even},\lambda_{odd})_C\longleftrightarrow (\rho'_{odd}, \lambda'_{even})_B.$$
However, this $S$-duality does not possess the commutativity shown in Fig.(\ref{ws}). Such a map leads to another class of problematic rigid surface operators \cite{rso}. These phenomena indirectly suggest that our proposed $S$-duality map (Proposition \ref{sm}) is more reasonable.

\paragraph{Weak Rigid Surface Operators}
\begin{figure}[!ht]
	\begin{center}
		\includegraphics[width=6in]{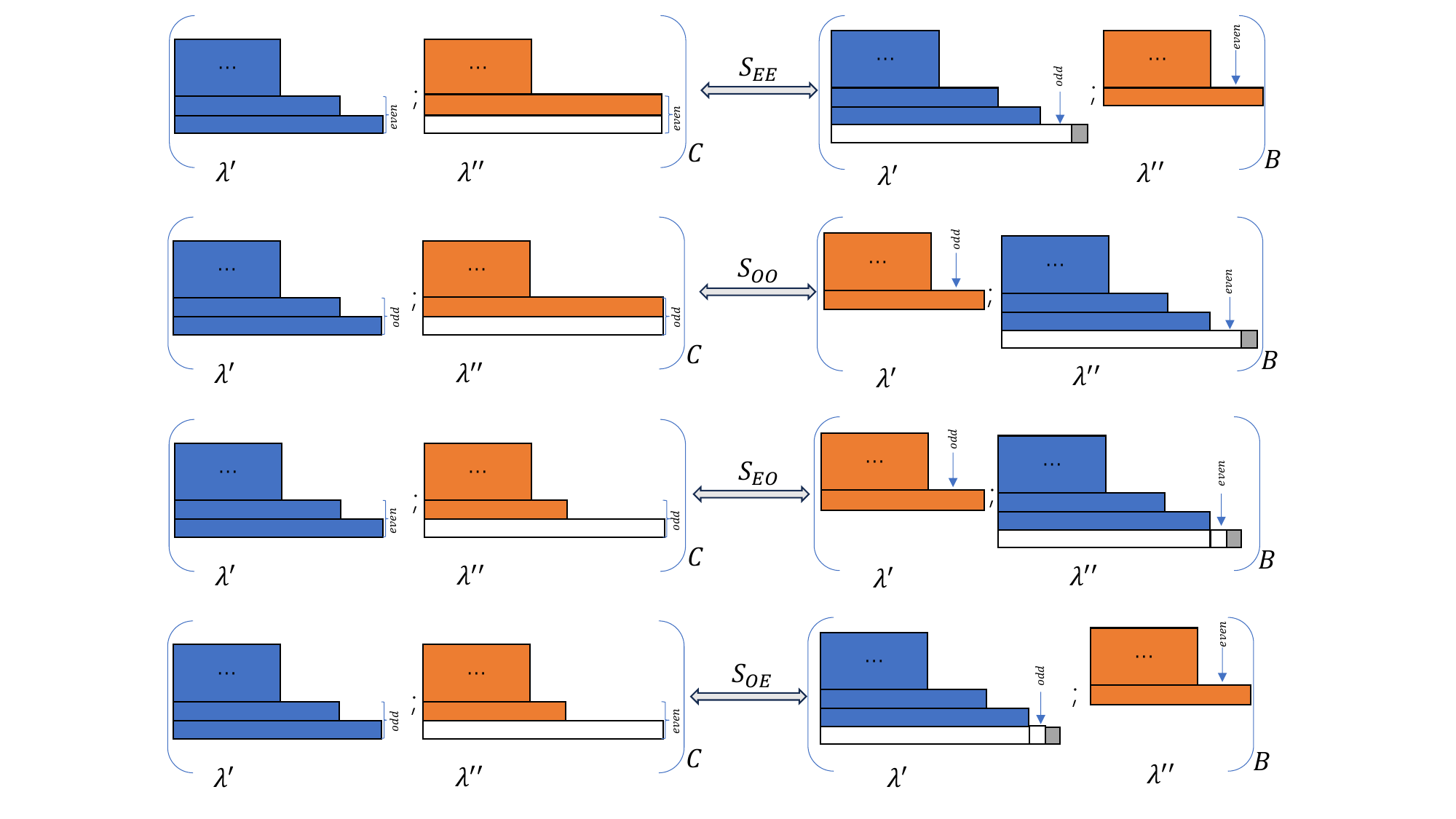}
	\end{center}
	\caption{Non-rigid surface operators generated by the $S$-duality map.}
	\label{nr}
\end{figure}

As can be seen from the commutative diagram in Fig.(\ref{nr}), the vast majority of rigid surface operators are mapped to rigid surface operators under duality. If a rigid surface operator maps to a non-rigid one, the violation of the rigid condition must follow one of two patterns:
\begin{itemize}
	\item Part 1 appears exactly twice only in the $B_n$ or $D_n$ theories. This occurs in the $S_{EO}$ and $S_{OE}$ maps.
	\item The gap strictly satisfies $\lambda_{l}-\lambda_{l-1}=2$ in the $C_n$ theory. This occurs in the $S_{EE}$ and $S_{OO}$ maps.
\end{itemize}

Both cases of non-rigid surface operators are related to the deformation of the abelian $SO(2)$ subgroup of the rigid surface operator \cite{GW08}.

In the first case, assume the non-rigid partition is $\lambda_1^{n_1}\lambda_2^{n_2}\cdots 2^{n_{m}} 1^2 $. The corresponding centralizer group is $S(O(2)^{n_m} \times O(2))$, where the notation $S(\cdot)$ denotes a double cover. Note that $n_m$ is even. We can perform a deformation on the Jordan form $diag(N,\cdots,N)$ corresponding to the parts $2^{n_m}1^2$, where $N=\left( \begin{array}{cc} 1 & 1 \\ 0 & 1 \end{array} \right)$. The deformation takes the form $I_2\cdots\otimes A \otimes\cdots I_2 \otimes A$, with $A=\left( \begin{array}{cc} a & b \\ c & d \end{array} \right) \in SO(2)$. This deformation reduces the centralizer but leaves the dimension unchanged. If the deformation breaks all discrete symmetries, the centralizer becomes $SO(2)^{n_m} \times SO(2)$. We term this type of non-rigid surface operator a \textit{weak rigid surface operator}, as discrete symmetries ensure they remain isolated orbits.

In the second case, assume the non-rigid partition is $\lambda_1^{n_1}\lambda_2^{n_2}\cdots 2^{n_{m}}$. Its corresponding commutator subgroup is $S( O(2)^{n_m})$. Note that the parity of $n_m$ is indefinite. Deforming the parts $2^{n_m}$ similarly breaks its discrete symmetry. This is also a class of weak rigid surface operators. However, when $n_m=1$, the centralizer can only be $SO(2)$, not its double cover. Lacking the protection of discrete symmetry, it is not an isolated orbit. This type of non-rigid surface operator may correspond to a special "limit point" or "singularity" in the parameter space.

This suggests that the explanation for the dual matching of rigid surface operators involves deeper quantum effects. Hopefully, our constructions will be helpful in making further progress.

\paragraph{Conclusions}
The simple and clear $S$-duality map we proposed can always find an $S$-dual pair for rigid surface operators, thereby resolving the mismatch problem associated with rigid surface operators. Under this map, all $S$-dual pairs previously identified by Gukov and Witten  can be reproduced \cite{sss} . Even in the special rigid class where our scheme differs from theirs, all invariants remain perfectly consistent. Comparatively, our proposal is more natural and aligns with existing mathematical results. The map possesses uniqueness under the requirement of Occam's Razor.

A similar approach can be applied to analyze $D_n$-type rigid surface operators. For the gauge group $SO(8)$, the results obtained are consistent with those in  \cite{sss}. However, the $D_n$ theory is self-dual, meaning the $S$-duality map is simultaneously a Type $I$ and Type $II$ map. When $N$ is large, $S$-duality cannot be realized simply by moving a single row. Therefore, a principle is needed to determine which Type $I$ map constitutes the true $S$-duality. In this context, the commutative diagram in Fig.(\ref{ws}) can serve as a starting point for constructing the $S$-duality map. We will discuss the case of $D_n$-type theories in more detail elsewhere.

Our results also contain subtle implications. Our proposed $S$-duality is completely determined by the symbol invariant and its construction. Thus, it would be interesting to rigorously prove the conjecture that exact $S$-duality implies discrete invariants. Our $S$-duality map establishes duality relations between all partition pairs on both sides of the dual theories, regardless of whether they satisfy rigid conditions. This extends beyond the scope of $S$-duality for rigid surface operators. The connection between conjugacy orbits of dual theories is itself a vast subject, with clear mathematical conclusions existing only for certain conjugacy classes \cite{l8}. Our results significantly refine the classification of conjugacy classes of dual groups and provide a novel physical perspective.

\section*{Acknowledgments}
We have benefited a lot from Professor Ke Wu’s detailed, sentence-by-sentence explanations of this topic many years ago.
We also thank Professor Pimo He from the Department of Physics, Zhejiang University, for his support.
Chuanzhong Li is supported by the National Natural Science Foundation of China (Grant No.12071237).

\section{Supplementary Materials }
In this section, we present additional examples of 
$S$-dual pairs from Gukov and Witten \cite{GW08}, which coincide with the results given in Proposition \ref{sm}.

\begin{table}[h!]
	\centering
	\small
	\begin{tabular}{lll:llllll}
	$B_{2}$ & $S$ & $C_{2}$ & $B_{2}$ & $C_{2}$ & Sym & FP & Dim \\
	\hline
	$(1;1^4)$ &  $S_{EO}$    & $(2\,1^{2};\emp)$ & $(1;1^4)$ & $(2\,1^{2};\emp)$ & $\left(\begin{array}{@{}c@{}c@{}c@{}c@{}} 1&&1 \\ &0& \end{array}\right)$ & $[1;1]$ & 4 \\
	$(2^{2}\,1;\emp)$ &  $S_{EE}$  & $(1^{2};1^{2})$ & $(2^{2}\,1;\emp)$ & $(1^{2};1^{2})$ & $\left(\begin{array}{@{}c@{}c@{}c@{}c@{}} 0&&0\\ &2& \end{array}\right)$ & $[2;\emp]$ & 4 \\
\end{tabular}
	\caption{$S$ dual pairs of rigid surface operators between $B_2$ and $C_2$ theories.}
\end{table}

\begin{table}[h!]
	\centering
	\small
	\begin{tabular}{lll:llllll}
	$D_{4}$ & $S$ & $D_{4}$ & $D_{4}$ & $D_{4}$ & Sym & FP & Dim \\
	\hline
	$(2^{2}\,1^{4};\emp)$ & I Type & $(2^{2}\,1^{4};\emp)$ & $(2^{2}\,1^{4};\emp)$ & $(2^{2}\,1^{4};\emp)$ & $\left(\begin{array}{@{}c@{}c@{}c@{}c@{}c@{}} 1&&1&&1\\ &0&&1& \end{array}\right)$ & $[2\,1^2;\emp]$ & 10 \\
	$(3\,2^{2}\,1;\emp)$ & I Type & $(3\,2^{2}\,1;\emp)$ & $(3\,2^{2}\,1;\emp)$ & $(3\,2^{2}\,1;\emp)$ & $\left(\begin{array}{@{}c@{}c@{}c@{}c@{}} 2&&2\\ &0& \end{array}\right)$ & $[\emp;1^4]$ & 16 \\
	$(1^{4};1^4)$ & I Type & $(1^{4};1^4)$ & $(1^{4};1^4)$ & $(1^{4};1^4)$ & $\left(\begin{array}{@{}c@{}c@{}c@{}c@{}} 2&&2\\ &0& \end{array}\right)$ & $[\emp;1^4]$ & 16 \\
\end{tabular}
	\caption{$S$ dual pairs of $D_4$ theory.}
\end{table}

\begin{table}[h!]
	\centering
	\small
	\begin{tabular}{lll:llllll}
	$B_{3}$ & $S$ & $C_{3}$ & $B_{3}$ & $C_{3}$ & Sym & FP & Dim \\
	\hline
	$(1;1^6)$ & $S_{EO}$ & $(2\,1^{4};\emp)$ & $(1;1^6)$ & $(2\,1^{4};\emp)$ & $\left(\begin{array}{@{}c@{}c@{}c@{}c@{}c@{}c@{}c@{}c@{}} 1&&1&&1 \\ &0&&0& \end{array}\right)$ & $[1^2;1]$ & 6 \\
	$(2^{2}\,1^{3};\emp)$ & $S_{EE}$ & $(1^{2};1^{4})$ & $(2^{2}\,1^{3};\emp)$ & $(1^{2};1^{4})$ & $\left(\begin{array}{@{}c@{}c@{}c@{}c@{}c@{}c@{}c@{}c@{}} 0&&0&&0\\ &1&&2& \end{array}\right)$ & $[2\,1;\emp]$ & 8 \\
	$(1^3;1^4)$ &  $S_{EO}$  & $(2^{3};\emp)$ & $(1^3;1^4)$ & $(2\,1^{2};1^2)$ & $\left(\begin{array}{@{}c@{}c@{}c@{}c@{}c@{}c@{}c@{}c@{}} 1&&1 \\ &1& \end{array}\right)$ & $[\emp;1^3]$ & 12 \\
	$(1;2^2\,1^2)$ &  $S_{EO}$  & $(2\,1^{2};1^2)$ & $(1^3;1^4)$ & $(2\,1^{2};1^2)$ & $\left(\begin{array}{@{}c@{}c@{}c@{}c@{}c@{}c@{}c@{}c@{}} 1&&1 \\ &1& \end{array}\right)$ & $[\emp;1^3]$ & 12 \\
\end{tabular}
	\caption{$S$ dual pairs of  rigid surface operators between $B_3$ and $C_3$ theories.}
\end{table}

\begin{table}[htbp]
	\centering
\renewcommand{\arraystretch}{1.3}
\begin{tabular}{llll:lllll	}
	\hline
	Num &
	SO(13) &
	S &
	Sp(12) &
	SO(13) &
	Sp(12) &
	Dim &
	Sym &
	FP \\
	\hline
	14 &
	$(2^2\,1;1^8)$ &  $S_{EO}$
	&
	$(3^2\,2\,1^4\,;\emp)$ &
	$(1^3;2^2\,1^6)$ &
	$(3^2\,2\,1^4\,;\emp)$ &
	44 &
	$\left(\begin{array}{@{}c@{}c@{}@{}c@{}c@{}c@{}c@{}c@{}c@{}}
		1&&1&&1&&1 \\
		&0&&0&&2&
	\end{array}\right)$ &
	$[3\,1^2;1]$ \\
	15 &
	$(1^3;2^2\,1^6)$ &  $S_{EO}$
	&
	$(2^3\,1^4\,;1^2)$ &
	$(2^2\,1;1^8)$ &
	$(2^3\,1^4\,;1^2)$ &
	44 &
	$\left(\begin{array}{@{}c@{}c@{}@{}c@{}c@{}c@{}c@{}c@{}c@{}}
		1&&1&&1&&1 \\
		&0&&0&&2&
	\end{array}\right)$ &
	$[3\,1^2;1]$ \\
	16 &
	$(1;3\,2^2\,1^5)$ &   $S_{OO}$
	&
	$(2\,1^6\,;2\,1^2)$ &
	$(1;3\,2^2\,1^5)$ &
	$(2\,1^6\,;2\,1^2)$ &
	44 &
	$\left(\begin{array}{@{}c@{}c@{}@{}c@{}c@{}c@{}c@{}c@{}c@{}}
		1&&1&&2&&2 \\
		&0&&0&&0&
	\end{array}\right)$ &
	$[2\,1^2;2]$ \\
	17 &
	$(1^5;2^2\,1^4)$ &  $S_{EO}$
	&
	$(2^3\,1^2\,;1^4)$ &
	$(1^5;2^2\,1^4)$ &
	$(2^3\,1^2\,;1^4)$ &
	50 &
	$\left(\begin{array}{@{}c@{}c@{}c@{}c@{}c@{}}
		1&&1&&1 \\
		&1&&2&
	\end{array}\right)$ &
	$[3;1^3]$ \\
	18 &
	$(2^2\,1^3;1^6)$ &  $S_{EO}$
	&
	$(3^2\,2^3;\emp)$ &
	$(2^2\,1^3;1^6)$ &
	$(3^2\,2^3;\emp)$&
	50 &
	$\left(\begin{array}{@{}c@{}c@{}@{}c@{}c@{}c@{}}
		1&&1&&1 \\
		&1&&2&
	\end{array}\right)$ &
	$[3;1^3]$ \\
	19 &
	$(2^4\,1;1^4)$ &  $S_{EE}$
	&
	$(2^4\,1^4;\emp)$ &
	$(2^4\,1;1^4)$ &
	$(2^4\,1^4;\emp)$ &
	52 &
	$\left(\begin{array}{@{}c@{}c@{}c@{}c@{}c@{}}
		0&&1&&1 \\
		&2&&2&
	\end{array}\right)$ &
	$[3^2;\emp]$ \\
	20 &
	$(1^5;3\,2^2\,1)$ &  $S_{EE}$
	&
	$(4\,3^2\,2;\emp)$ &
	$(1^5;3\,2^2\,1)$ &
	$(4\,3^2\,2;\emp)$ &
	56 &
	$\left(\begin{array}{@{}c@{}c@{}@{}c@{}c@{}c@{}c@{}}
		0&&2&&2 \\
		&1&&1&
	\end{array}\right)$ &
	$[3;2\,1]$ \\
\end{tabular}
	\caption{$S$ dual pairs of  rigid surface operators between $B_6$ and $C_6$ theories.}
\end{table}

\end{document}